\documentclass[times, twoside, watermark]{zHenriquesLab-StyleBioRxiv}
\usepackage{blindtext}
\usepackage{lipsum}
\leadauthor{Percannella} 

\begin{document}

\title{Mitosis detection in domain shift scenarios: a Mamba-based approach}
\shorttitle{Approach for MIDOG 2025}

\author[1]{Gennaro Percannella}
\author[1]{Mattia Sarno}
\author[1]{Francesco Tortorella}
\author[1]{Mario Vento}
\affil[1]{Department of Information and Electrical Engineering and Applied Mathematics (DIEM), University of Salerno, Via Giovanni Paolo II
132, Fisciano, 84084, Salerno, Italy}

\maketitle

\begin{abstract}
Mitosis detection in histopathology images plays a key role in tumor assessment. Although machine learning algorithms could be exploited for aiding physicians in 
accurately performing such a task, these algorithms suffer from significative performance drop when evaluated on images coming from domains that are different from the training ones.
In this work, we propose a Mamba-based approach for mitosis detection under domain shift, inspired by the promising performance demonstrated by Mamba in medical imaging
segmentation tasks. Specifically, our approach exploits a VM-UNet architecture for carrying out the addressed task, as well as stain augmentation operations 
for further improving model robustness against domain shift. Our approach has been submitted to the track 1 of the \textbf{MI}tosis \textbf{DO}main \textbf{G}eneralization (MIDOG) challenge. Preliminary
experiments, conducted on the MIDOG++ dataset, show large room for improvement for the proposed method.
\end {abstract}

\begin{keywords}
Mitosis detection, Domain shift, Mamba 
\end{keywords}

\begin{corrauthor}
masarno@unisa.it
\end{corrauthor}

\section*{Introduction}
Histopathology plays a crucial role in tumor characterization, allowing the acquisition of images at high resolution, that can be used by pathologists for analysing
tumor traits, to determine tumor stage. Among such traits, mitotic count, i.e., the assessment of cells
undergoing cell division (mitosis) in a defined area (typically $2 mm^2$) \cite{AUBREVILLE2023102699}, 
represents a general index of the tumor stage. Detecting mitoses in histopathology images, however, is a complex and time-consuming operation,
due to several factors: first, there is a high intraclass variability in mitotic figures, due to the large variability in the morphology of mitotic
nuclei; moreover, there exist non-mitotic cells that show high similarity with mitotic ones, which may lead to false detections.\\
Machine learning models have been proposed in several works for automatically carrying out such a task \cite{wang2023generalizable}\cite{jahanifar2024mitosis}. However, these models suffer from significative performance drop
when evaluated on images belonging to domains that are different from the training ones. In particular, in histopathology, domains may vary due to factors related to the
image acquisition scanners or protocols, due to the analysed tissue or tumor type, as well as due to the species which subjects belong to.\\
The Mitosis Domain Generalization (MIDOG) challenge \cite{ammeling_mitosis_2025} aims at promoting the development of machine learning models for mitosis detection and characterization, robust to such shifts.
In this work, we present our approach for mitosis detection, submitted to the track 1 of the MIDOG25 challenge. Specifically, our approach traces the mitosis detection problem back to 
a mitosis segmentation one, in line with previous approaches \cite{wang2023generalizable}\cite{jahanifar2024mitosis}, and performs this task through a Mamba-based \cite{liu2024vmamba} architecture, which showed promising performance in several medical image segmentation 
tasks in domain shift scenarios \cite{cheng2025mamba}. To further improve model generalization capability against domain shift, the proposed method
exploits style augmentation operations, by applying stain perturbations to the training images.

\section*{Material and Methods}

\subsection*{Mitosis segmentation mask extraction}
In order to trace the addressed mitosis detection problem back to a mitosis segmentation one, a ground truth segmentation mask for each mitosis has been extracted, by means of NuClick \cite{koohbanani2020nuclick}, 
an open-source segmentation model that performs nuclei segmentation in histopathology images, based on point annotations related to such nuclei. In this work, the center of each mitosis has been
extracted from the metadata of the provided dataset, and it was passed to NuClick, for the extraction of the segmentation map related to that mitosis.

\subsection*{Network architecture}
\begin{figure*}
  \centering
  \includegraphics[width=1.5\columnwidth]{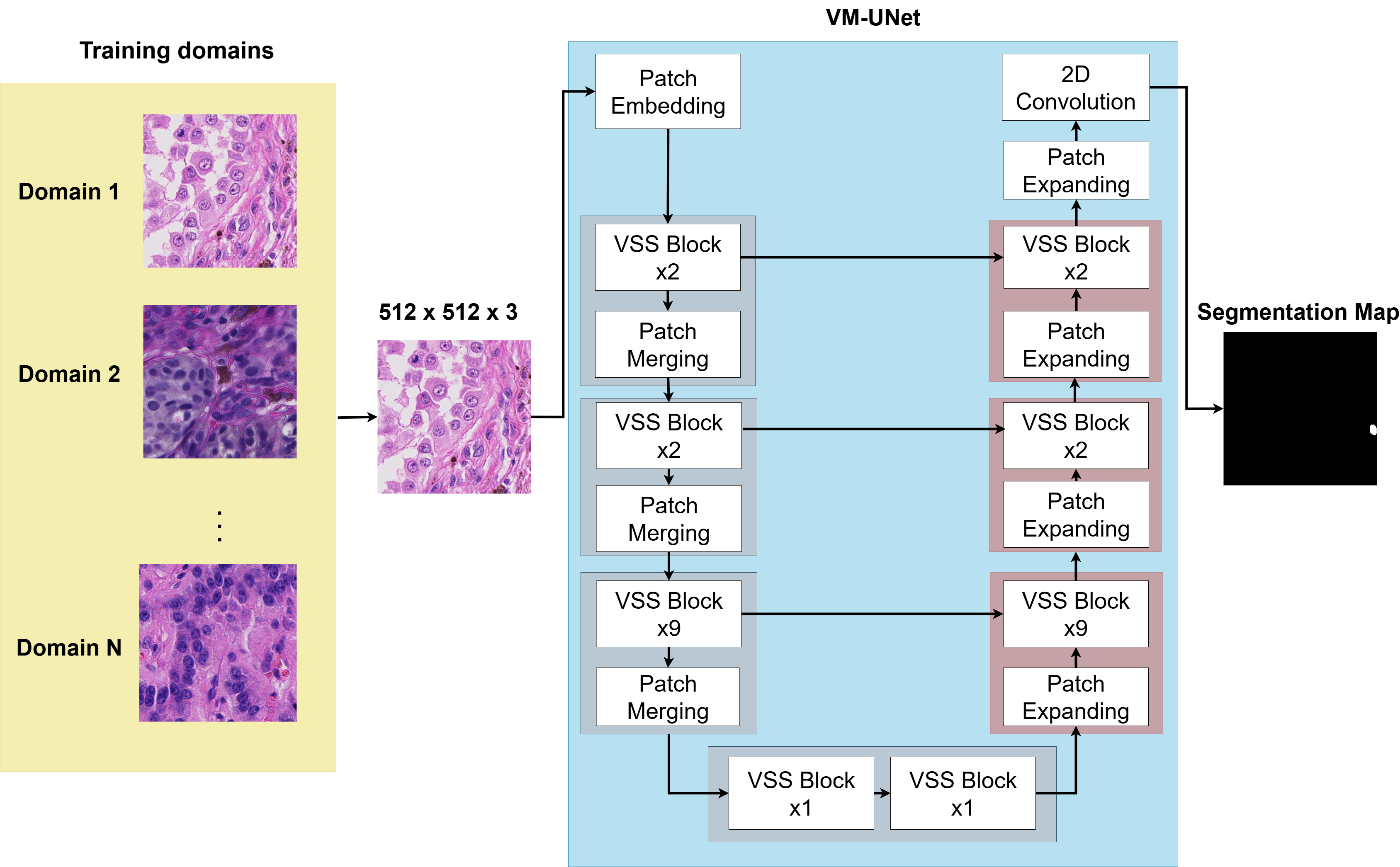}
  \caption{VM-UNet architecture}
  \label{fig:vm}
\end{figure*}

The model exploited by the proposed approach is a VM-UNet \cite{ruan2024vm} pretrained on ImageNet \cite{fei2009imagenet}, shown in Figure \ref{fig:vm}. This network consists of a Patch-Embedding layer, followed by an encoder-decoder architecture with skip connections. The Patch-Embedding layer, in particular,
divides the input image into non-overlapping patches of size $4\times4$, and subsequently projects each patch into a vector of 96 features, generating an embedded image of size $\frac{H}{4} \times \frac{W}{4} \times 96$, that is passed to the encoder.\\
As regards the encoder structure, instead, it consists of 4 sequential stages, each of which exploits Mamba VSS blocks \cite{liu2024vmamba} for feature extraction. At the end of the first three stages, patch merging blocks are further exploited for reducing the spatial dimensions
of the input image by a factor of 4, while doubling the number of channels.\\ 
Similar to the encoder, the network decoder consists of 4 sequential stages: at the beginning of the first three stages, the spatial dimensions of the analysed feature maps is increased by a factor
of 4, thanks to patch expanding blocks. Additionally, VSS blocks are exploited for decreasing the number of channels of such feature maps. 
\\After the final decoder stage, a projection layer restores the received feature maps to the size of the input image, to match the size of
segmentation target.

\subsection*{Style augmentation}
To improve model robustness against domain shift, style augmentation has been performed, by applying random stain pertubation to the training images: specifically, first, the stain $S$ and the concentration $C$ matrices of each image $I_0$ are estimated,
through the $Vahadane$ method \cite{vahadane2016structure}. Then, scaling and shifting operations with random coefficients $\alpha$ and $\beta$ are applied to the stain matrix and on the concentration matrix respectively, to obtain new altered matrices. These altered matrices are finally converted back to the RGB space, as reported in Equation \ref{eq:intensita}, 
generating the stain perturbed image $I$.

\begin{equation}
I = I_0 \exp\left(-S(\alpha C + \beta)\right)
\label{eq:intensita}
\end{equation}

\subsection*{Model training}
The proposed network was trained on tiles of size $512 \times 512$, extracted from the training images with a $80\%$ of overlap. On such tiles, the model was trained for 100 epochs, using AdamW as optimizer with learning rate $5 \times 10^{-4}$, and batch size of 24. To handle the severe class imbalance that characterizes the addressed problem, each batch was built to have the same number of positive tiles, i.e., tiles containing at least one mitosis, and negative ones, i.e., tiles that do not contain any mitosis. The combination between Dice loss \cite{milletari2016v} and Focal loss \cite{lin2017focal} was finally exploited as a loss function.

\subsection*{Inference}
At inference time, each analysed image is divided into overlapping tiles of size $512 \times 512$, with an overlap level of $80\%$, and model predictions on such tiles are aggregated to generate the predicted segmentation mask associated to the input image.\\
The generated segmentation map is then post-processed, to get the coordinates of the detected mitoses: in particular, first, morphological dilation is performed on the predicted segmentation mask, to avoid that the daughter cells generated from a mitotic event, which appear more clear in the late phases of a mitosis, are detected by the proposed method as two separate mitotic figures. Then, from the processed mask, 
the coordinates of the centers of the bounding boxes of minimum area enclosing each connected component are extracted, and returned as the locations of the detected mitoses. 
\\
To further improve robustness of the proposed method against domain shift, model ensembling is exploited at inference time, by aggregating the predictions of the best found models in the performed cross-validation experiments.
\section*{Results}

\subsection*{Dataset}
The proposed approach was evaluated on the MIDOG++ dataset \cite{aubreville_comprehensive_2023} which contains mitosis annotations for 11,937 mitotic figures across 503 individual tumor cases, spanning 7 different domains. Each domain, in particular, is characterized by images referring to a specific tumor type, and acquired through a specific scanner on subjects belonging to a certain
species.
\subsection*{Experimental protocol}
Experiments have been carried out according to a leave-one-domain-out experimental protocol, over the domains that characterize the MIDOG++ dataset. Specifically, in each experiment, 6 domains have been exploited for model training, while the remaining one was used for validation. Model selection was performed based on the value of the loss function on the validation domain.
\\
In the conducted experiments, model performance have been assessed by evaluating the F1-score on the validation domains of the performed leave-one-domain-out.

\subsection*{Experimental results}
In Table \ref{tab:table exp} results of the conducted preliminary experiments on the MIDOG++ dataset are shown. As it's possible to observe, a Mamba-based U-Net has higher generalization capability, compared to a standard convolutional U-Net \cite{ronneberger2015u}. Furthermore, the use of stain augmentation
positively contributes to its generalization capability on domains unseen at training time. In spite of that, there is still much room for improvement.\\
On the preliminary test set of the track 1 of the MIDOG25 challenge, the proposed approach achieves a F1-score of 0.759.

\begin{table}[h!]
\centering
\footnotesize
\begin{tabular}{|c|c|}  
\hline
\textbf{Model} & \textbf{Validation domain}  \\
\hline
\textbf{U-Net \cite{ronneberger2015u}} & $0.656_{\pm 0.094}$  \\
\hline
\textbf{VM-UNet} & $0.710_{\pm 0.073}$ \\
\hline
\textbf{VM-UNet + Stain Aug} & $\textbf{0.736}_{\pm 0.063}$ \\
\hline
\end{tabular}
\caption{Mean and standard deviation of the F1-score on the validation domain across the leave-one-domain-out experiments}
\label{tab:table exp}
\end{table}

\section*{Discussion}
In this work, a Mamba-based approach for mitosis detection under domain shift has been proposed and submitted to the track 1 of the MIDOG25 challenge. The proposed approach traces the mitosis detection problem back to a mitosis segmentation one and exploits a VM-UNet and stain augmentation for dealing with it. Although in the conducted preliminary experiments the proposed
approach shows encouraging performance, there is still much room for improvement.

\section*{Bibliography}
\bibliography{literature}

\end{document}